\documentclass[prb,a4paper,twocolumn,showpacs]{revtex4-1}	

\usepackage{graphicx}
\usepackage{epsfig}
\usepackage{amsmath}
\usepackage{epstopdf}
\usepackage{subfigure}
\usepackage{color}

\begin{document}

\title{Plasmons in anisotropic Dirac systems}

\author{Roland Hayn}
\email{roland.hayn@im2np.fr}
\affiliation{Aix-Marseille Univ., CNRS, IM2NP-UMR 7334, 13397 Marseille Cedex 20,
France}
\affiliation{Leibniz Institute for Solid State and Materials Research IFW Dresden, Helmholtzstr. 20, 01069 Dresden, Germany}

\author{Te Wei}
\affiliation{Aix-Marseille Univ., CNRS, IM2NP-UMR 7334, 13397 Marseille Cedex 20,
France}

\author{Vyacheslav M.  Silkin}
\affiliation{Donostia International Physics Center (DIPC), 20018 San Sebasti\'an/Donostia, Basque Country, Spain}
\affiliation{ Departamento de Pol\'{\i}meros y Materiales Avanzados: F\'{\i}sica, Qu\'{\i}mica y Tecnolog\'{\i}a,
Facultad de Ciencias Qu\'{\i}micas, Universidad del Pa\'{\i}s Vasco UPV/EHU, 20080 San Sebasti\'an/Donostia, Basque Country, Spain}
\affiliation{IKERBASQUE, Basque Foundation for Science, 48013 Bilbao, Basque Country, Spain}

\author{Jeroen van den Brink}
\affiliation{Leibniz Institute for Solid State and Materials Research IFW Dresden, Helmholtzstr. 20, 01069 Dresden, Germany}
\affiliation{Institut f\"ur Theoretische Physik and W\"urzburg-Dresden Cluster of Excellence ct.qmat, Technische Universit\"at Dresden, 01062 Dresden, Germany}

\date{\today}

\begin{abstract}
We consider the plasmon excitations in anisotropic two-dimensional Dirac systems, be it either anisotropic graphene or surfaces of topological insulators.
Generalizing the exact density-density response function one finds a plasmon dispersion that is anisotropic already at the lowest frequencies. Asymptotic expressions are obtained for the dispersion in this regime.
We show that the plasmon properties of the complete material class of anisotropic Dirac systems are characterized by just two dimensionless material parameters. The strong anisotropy can be used to guide the plasmon modes, introducing new functionalities to the field of Dirac plasmonics.

\end{abstract}

\maketitle

\section{Introduction}

Graphene and topological insulators (TI) are two-dimensional (2D) Dirac systems \cite{Geim07, Kane05}
in the sense that they have a linear electron (and hole) dispersion and a Dirac point where the Fermi surface shrinks to zero. The peculiarities of relativistic
electrons and the high Fermi velocity make them unique systems to study fundamental phenomena like spin-momentum
locking and open many interesting applications in nano-electronics. Replacing the spin in TI by the pseudo-spin in graphene
leads to a high formal analogy between both types of systems,
be it that the number of Dirac cones that are present in the 2D Brioullin zone in one case is odd and in the other even.
In the doped case, these Dirac systems allow for
collective charge excitations  -- plasmons -- that are different from both bulk and surface plasmons of ordinary metals.
A pure 2D Dirac plasmon, like its 3D counterpart, has no direct coupling to light due to the momentum mismatch.
However, such a
coupling can be created by proper surface modification that break translation symmetry, for instance by grating or nano-structuration. This allows for interesting applications such as terahertz photodetectors, motivating the field of graphene plasmonics, or more in general, Dirac plasmonics.\cite{kochnl11,Grigorenko12,diornn13,gaap14}

Here we concentrate on systems having an anisotropic Dirac cone in particular with a high  factor of anisotropy  $A=v_x/v_y$ between two extremal Fermi velocities in the two perpendicular directions $x$ and $y$.
A large factor of $A=18$ was for instance predicted for the topological surface states of the 3D TI HgS, \cite{Virot11} but other TI's can have large anisotropy factors as well. \cite{Zhang11} Experimentally anisotropic Dirac cones were detected recently by angle resolved photoemission in for instance
Ru$_2$Sn$_3$,\cite{gievsr14} CaMnBi$_2$,\cite{fewasr14}  BaMnBi$_2$, and BaZnBi$_2$.\cite{Ryu18}
In graphene the Dirac cone warping produces some anisotropy in the dispersion of the 2D and acoustic plasmons.\cite{himiepl09,gayussc11,denoprb13,pisinjp14}
External strain can cause spatial anisotropy in graphene, but the expected changes in the plasmon anisotropy are rather small.\cite{Choi10}

Quite a considerable amount of theoretical work had been devoted to tilted Dirac cones which can be found
in $\alpha$-(BEDT-TTF)$_2$I$_3$ (BEDT-TTF=bis(ethylene-dithio)tetrathiafuva)
under pressure,\cite{Tajima06} in some other organic quasi-two-dimensional materials as well as in orthorombic borophene.\cite{Feng17}
The analytical result for the imaginary part of the density-density response has been given in
Ref.\ \onlinecite{Nishine10} and for the real part in Ref.\  \onlinecite{Sadhukhan17}. There, also a slight anisotropy
was included. Plasmons of a tilted cone in a magnetic field were analyzed in Ref.\ \onlinecite{Sari14}.
However, the analytical formula of Ref.\ \onlinecite{Sadhukhan17}
was criticized in Ref.\ \onlinecite{Jalali18} and we will clarify that point here for any possible anisotropic Dirac system.
We will not consider the effect of tilting, but rather only spatial anisotropy that lowers in-plane rotation symmetry which is the usual case for  anisotropic TI's and for this situation will provide analytical expressions for the full plasmon dispersion and certain limiting cases.
We are going to derive handy analytical formulas for the anisotropic plasmon dispersion of a general anisotropic
Dirac system being characterized by just 2 dimensionless material parameters.

\section{Hamiltonian and charge response}

We are considering electrons confined to two dimensions with Coulomb interactions.
The Hamiltonian of an anisotropic Dirac system is given by
\begin{equation}
\label{eq1}
H=\sum_{\bf k} \varepsilon_{\bf k} c_{{\bf k} \uparrow}^{\dagger} c_{{\bf k} \downarrow} +
\varepsilon_{\bf k}^{*} c_{{\bf k} \downarrow}^{\dagger}  c_{{\bf k} \uparrow} \; ,
\end{equation}
where $c^\dagger$/$c$ represent fermion creation/annihilation operators, $\bf k$ the 2D wavevector and the energy is given in terms of the velocities $v_x$/$v_y$ in $x$/$y$ direction as
\begin{equation}
\label{eq2}
\varepsilon_{\bf k} = v_y k_y +  {\rm i} v_x k_x
= \lvert \varepsilon_{\bf k} \rvert \exp ( {\rm i} \Phi_{\bf k}) \; .
\end{equation}
The Hamiltonian describes anisotropic topological insulators
or graphene if one replaces spin by pseudo spin
and adds valley and spin degeneracies.
The
plasmon dispersion can then be obtained by calculating the dielectric function in random phase approximation (RPA). The
dielectric function at 2D wave vector ${\bf q}$ and energy transfer $\omega$ is related with the charge susceptibility (or density-density response function)
\begin{equation}
\label{eq3}
\chi (\bf{q},\omega)=\langle \langle \rho_{\bf q} ; \rho_{\bf -q} \rangle \rangle
\end{equation}
with
\begin{equation}
\rho_{\bf q}=\sum_{{\bf k} \sigma} c_{{\bf k} \sigma}^{\dagger} c_{{\bf k+q} \sigma}
\nonumber
\end{equation}
being expressed via a retarded Green's function. In RPA we obtain
\begin{equation}
\chi ({\bf q},\omega)=\frac{\chi_0 ({\bf q},\omega)}{1-V({\bf q}) \chi_0 ({\bf q},\omega)} \; ,
\label{eq4}
\end{equation}
where $\chi_0$ is the electron-hole bubble (in graphical representation) and $V({\bf q})=e^2/(2 \lvert {\bf q} \rvert
\varepsilon_0 \varepsilon_{rel})$ is the Coulomb interaction in the 2D system.
Following the calculation for the isotropic case \cite{Wunsch06,Hwang07,Principi09} we generalize it to the anisotropic situation.
By diagonalizing (\ref{eq1})  one finds two energy branches
$\pm \lvert \varepsilon_{\bf k} \rvert = \lambda \lvert \varepsilon_{\bf k} \rvert$.
The unitary transformation
\begin{equation}
\tilde{c}_{{\bf k} \pm}
= (c_{{\bf k} \uparrow} \exp (- {\rm i} \phi_{\bf k} / 2) \pm c_{{\bf k} \downarrow} \exp ( {\rm i} \phi_{\bf k} / 2))/ \sqrt{2}
\nonumber
\end{equation}
diagonalizes the Hamiltonian (\ref{eq1}) and gives the
zero-order susceptibility as:
\begin{equation}
\nonumber
\chi_0 ({\bf q},\omega)=
\sum_{\lambda \lambda^{\prime}} \chi_0^{\lambda \lambda^{\prime}}=
g \sum_{{\bf k} \lambda \lambda^{\prime}}
\frac{F^{\lambda \lambda^{\prime}} (n_{{\bf k} \lambda} - n_{{\bf k+q} \lambda^{\prime}} ) }
{\omega + {\rm i} 0^+ + \lambda \lvert \varepsilon_ {\bf k} \rvert - \lambda^{\prime} \lvert \varepsilon_ {\bf k+q} \rvert} \; ,
\end{equation}
where $g$ is an eventual degeneracy ($g=4$ in graphene due to spin and valley degeneracy).
Also,
$\{ \lambda, \lambda^{\prime}\}=\pm 1$ denote the two branches of the dispersion and
$n_{{\bf k} \lambda}$ is in general the Fermi function
$n_{{\bf k} \lambda}=f(\lambda \lvert \varepsilon_{\bf k} \rvert - \varepsilon_F)$ of the $\lambda$ branch but we
restrict ourselves
here to zero temperature and $\varepsilon_F$ is the Fermi energy.
The form factor is
\begin{equation}
\nonumber
F^{\lambda \lambda^{\prime}}=(1 + \lambda \lambda^{\prime} \cos (\Phi_{\bf k+q} - \Phi_{\bf k}))/2 \; .
\end{equation}
We consider now a doped situation with a Fermi energy $\varepsilon_F$ lying in the positive branch $\lambda=+1$.
Since the negative branch is completely filled,  $\chi_0^{--}$ is zero. We are interested in the real part of
$\chi_0$ to determine the plasmon dispersion via the zero of the denominator of (\ref{eq4}). As in the isotropic case,
the plasmon dispersion is dominated by $\chi_0^{++}$ which can be expressed as:
\begin{equation}
\chi_0^{++}= g \iint_{\varepsilon_{\bf k} < \varepsilon_F}
\frac{ \rm d k_x \rm d k_y}{4 \pi^2} \frac{F^{++} (\lvert \varepsilon_{\bf k+q} \rvert - \lvert \varepsilon_{\bf k} \rvert  )}
{\omega^2 - (\lvert \varepsilon_{\bf k+q} \rvert - \lvert \varepsilon_{\bf k} \rvert  )^2} \; .
\label{eq5}
\end{equation}
After introducing vectors ${\bf K}$ and ${\bf Q}$ with $K_i=k_i v_i / v$,
$Q_i=q_i v_i / v$ ($i=\{ x,y \}$)
and $v^2=v_x v_y$ we can write
$\lvert \varepsilon_{\bf k} \rvert = v \lvert {\bf K} \rvert $ and cast integral (\ref{eq5}) into the same form as for the
isotropic case
\begin{equation}
\nonumber
\chi_0^{++}= g \iint_{\lvert {\bf K} \rvert <  \frac{\varepsilon_{F}}{v} }
\frac{ \rm d K_x \rm d K_y}{4 \pi^2} \frac{F^{++} v (\lvert {\bf K+Q} \rvert - \lvert {\bf K} \rvert  )}
{\omega^2 - v^2 (\lvert {\bf K+Q} \rvert - {\bf K} \rvert  )^2} \; .
\end{equation}
We also see that $\Phi_{{\bf k+q}} -\Phi_{{\bf k}}$ equals the angle between ${\bf K+Q}$ and ${\bf K}$. Therefore,
we can use for $\chi_0^{++}$ at wave vector ${\bf q}$ in the anisotropic case the expression for the isotropic case
$\chi_0^{++,{\rm iso}}$
at
wave vector ${\bf Q}$ which is also true for the other contributions $\chi_0^{+-}$ and $\chi_0^{-+}$. We find finally
\begin{equation}
\chi_0 ({\bf q},\omega)=\chi_0^{{\rm iso}} ({\bf Q},\omega) \; ,
\end{equation}
where we have to use the Fermi velocity $v=\sqrt{v_x v_y}$ in $\chi_0^{{\rm iso}}$.
The exact expression of $\chi_0^{{\rm iso}}$ in the isotropic case is well known, \cite{Wunsch06,Hwang07}
but it now depends on $Q = \lvert {\bf Q} \rvert$ instead of $q = \lvert {\bf q} \rvert$. The dependence
on the angle $\alpha$ of the plasmon propagation, where $q_x= q \cos \alpha$ and $q_y=q \sin \alpha$
can be cast into a directional factor $D$:
\begin{equation}
Q=q D \quad , \quad D= \sqrt{A  \cos^2 \alpha + \frac{\sin^2 \alpha}{A}} \; .
\label{eq14}
\end{equation}
Using the known expression for $\chi_0^{{\rm iso}}$, we find the exact expression for the density-density response
function in the anisotropic case. It can be expressed like
\begin{equation}
\chi_0({\bf q},\omega)=C g(\kappa,\nu) \quad , \quad
C=\frac{g \varepsilon_F}{2 \pi v^2 \hbar^2} \; ,
\label{eq8}
\end{equation}
in dependence on the dimensionless parameters
\begin{equation}
\nu=\frac{\hbar \omega}{\varepsilon_F} \quad , \quad \kappa=\frac{q}{q_F} D \; ,
\end{equation}
where we introduce $\hbar$ from now on with $q_F$ being an averaged Fermi wave vector defined
by $\varepsilon_F=\hbar v q_F$, and where
\begin{equation}
\nonumber
g(\kappa,\nu)=-1+ f \left( G_+ \left( \frac{2+\nu}{\kappa} \right) -
G_+ \left( \frac{2-\nu}{\kappa} \right) \right)
\end{equation}
and
\begin{equation}
\nonumber
f=\frac{\kappa^2}{8 \sqrt{\nu^2-\kappa^2}} \, , \,
G_+(x)=x\sqrt{x^2-1}-\ln{(x+\sqrt{x^2+1})} \; .
\end{equation}
This expression for the real part outside the continuum
of electron-hole excitations whose border is given by  $\nu=\kappa$ and $\nu=2-\kappa$
and where the imaginary part of $\chi_0$ is zero derives from
the complete expression given in Refs.\ \onlinecite{ Wunsch06,Hwang07}.
The analytical expression (\ref{eq8}) can also be obtained from the tilted case \cite{Sadhukhan17, Jalali18}
by putting the tilt angle to zero in which case the difference between Refs.\
\onlinecite{Sadhukhan17} and \onlinecite{Jalali18} disappears.

\begin{figure}
\centering
\includegraphics[width=0.9 \linewidth]{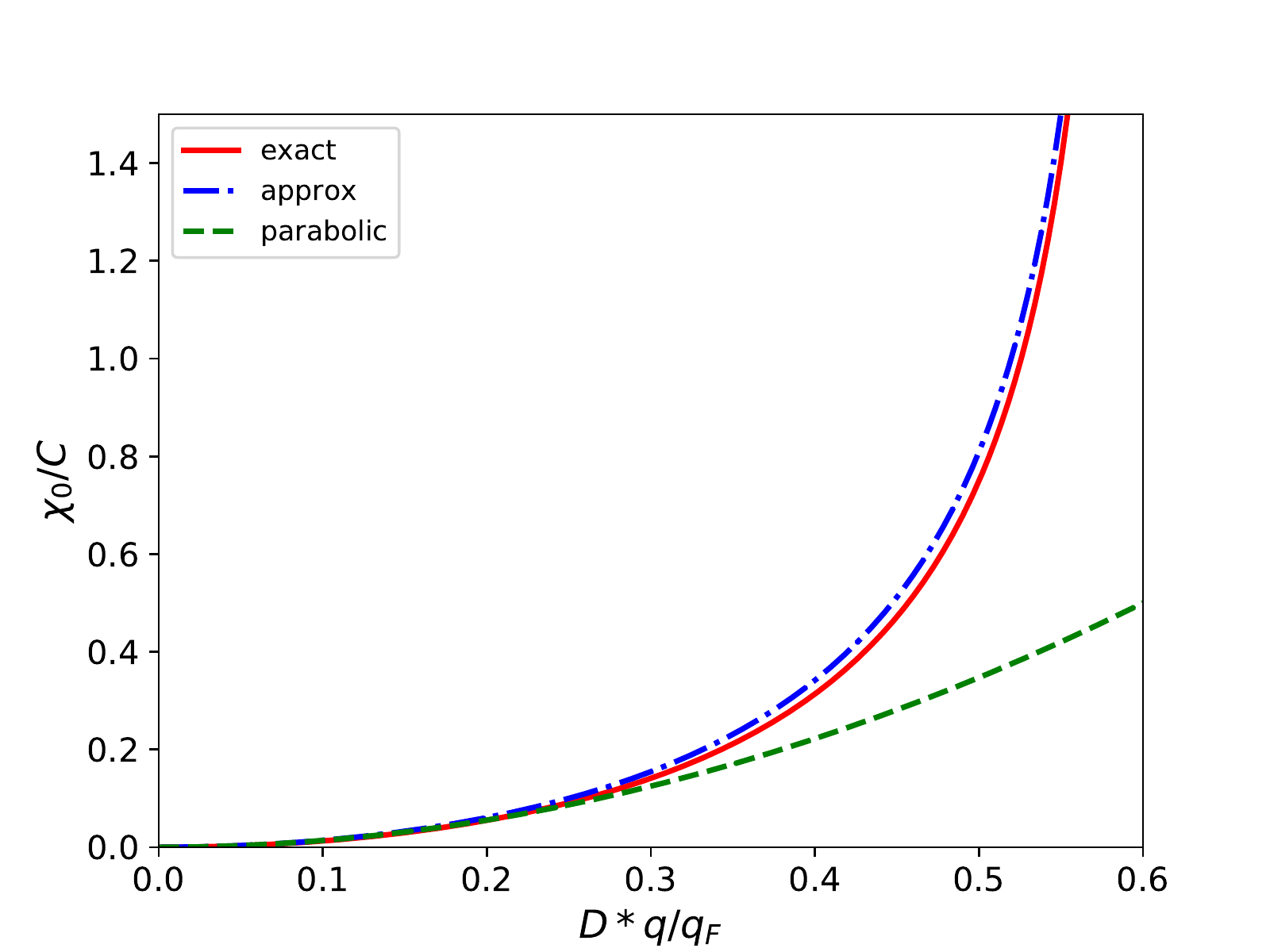}
\caption
{Real part of the zero-order density-density response function $\chi_0/C$ for $\nu=\hbar \omega / \varepsilon_F=0.6$.
Compared are the exact expression (red, full line) with the approximative one (Eq.\ (\ref{eq11}), blue dash-dotted line) and the
parabolic approximation (Eq.\ (\ref{eq12}), green dashed line).}
\label{f1}
\end{figure}

We are interested in the plasmon dispersion in the
hydrodynamic limit $\omega \to 0$ and $ q \to 0$ where we can use the
leading-order expression \cite{Raghu10}:
\begin{equation}
\label{eq11}
g(\kappa,\nu)=\frac{\nu^2}{\sqrt{\nu^2-\kappa^2}}-1 \; .
\end{equation}
For $\nu \gg \kappa$ that simplifies to
\begin{equation}
\label{eq12}
g(\kappa,\nu)=\frac{\kappa^2}{2 \nu^2} \; .
\end{equation}
To illustrate the different approximations we present them in Fig.\ \ref{f1} together with the exact expression for $\nu=0.6$.
At small $q$ all three expressions coincide, but $\chi_0$ diverges if $\kappa=D q / q_F$ approaches the
continuum of particle-hole excitations $\kappa=\nu$ which is not the case in the parabolic approximation in Eq.\  (\ref{eq12}).

\section{Plasmon}

The plasmon dispersion is determined by solving $V({\bf q}) \chi_0({\bf q},\omega)=1$ which is in dimensionless
form
\begin{equation}
\label{eq13}
\frac{q}{q_F}=2 \beta g(\kappa,\nu) \; ,
\end{equation}
where we introduce the dimensionless material parameter
\begin{equation}
\beta=\frac{g e^2}{8 \pi \varepsilon_0 \varepsilon_{rel} \hbar v} \; .
\end{equation}
In the parabolic approximation for small $q$ and $\omega$ the plasmon dispersion can be explicitly given,
\begin{equation}
\frac{\hbar \omega}{\varepsilon_F}=\sqrt{\beta} \sqrt{\frac{q}{q_F}} D \; ,
\label{eq15}
\end{equation}
and is especially simple. The square-root dispersion is of course characteristic to 2D systems.

Any anisotropic Dirac system is characterized by the degeneracy $g$, the Fermi velocity $v$, the anisotropy
$A=v_x/v_y$, the relative dielectric constant $\varepsilon_{rel}$,
and the Fermi energy $\varepsilon_F$ closely related with the filling of the Dirac cone.
The plasmon dispersion which is given by the solution of (\ref{eq13}) is
valid for any anisotropic Dirac system and
characterized by just two
material parameters $\beta$ and $A$.
At the same time, without tilting, the analytical result  (\ref{eq13}) is rather simple.


\begin{figure}
\centering
\includegraphics[width=1.0  \linewidth]{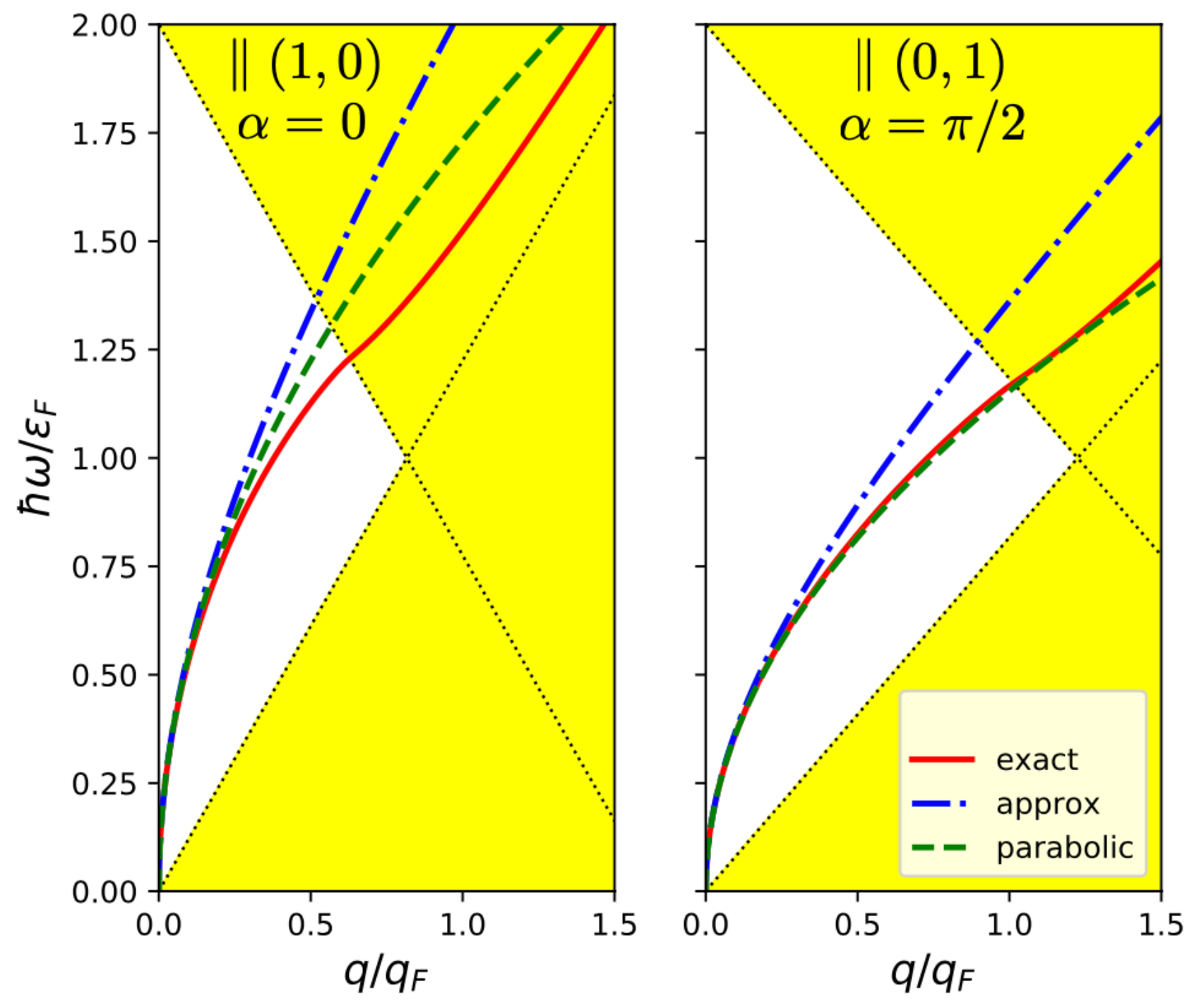}
\caption
{Plasmon dispersion for material parameters $\beta=2.0$ and $A=1.5$ for the two extremal plasmon propagation
directions
$\alpha=0$ (left) and $\alpha=\pi/2$ (right) using the exact expression or the two approximate
ones (see Fig.\ \ref{f1}) together with the corresponding boundaries
of the continuum of electron-hole excitations (yellow) shown by dotted lines.
}
\label{f2}
\end{figure}

\begin{figure}
\centering
\includegraphics[width=1.0 \linewidth]{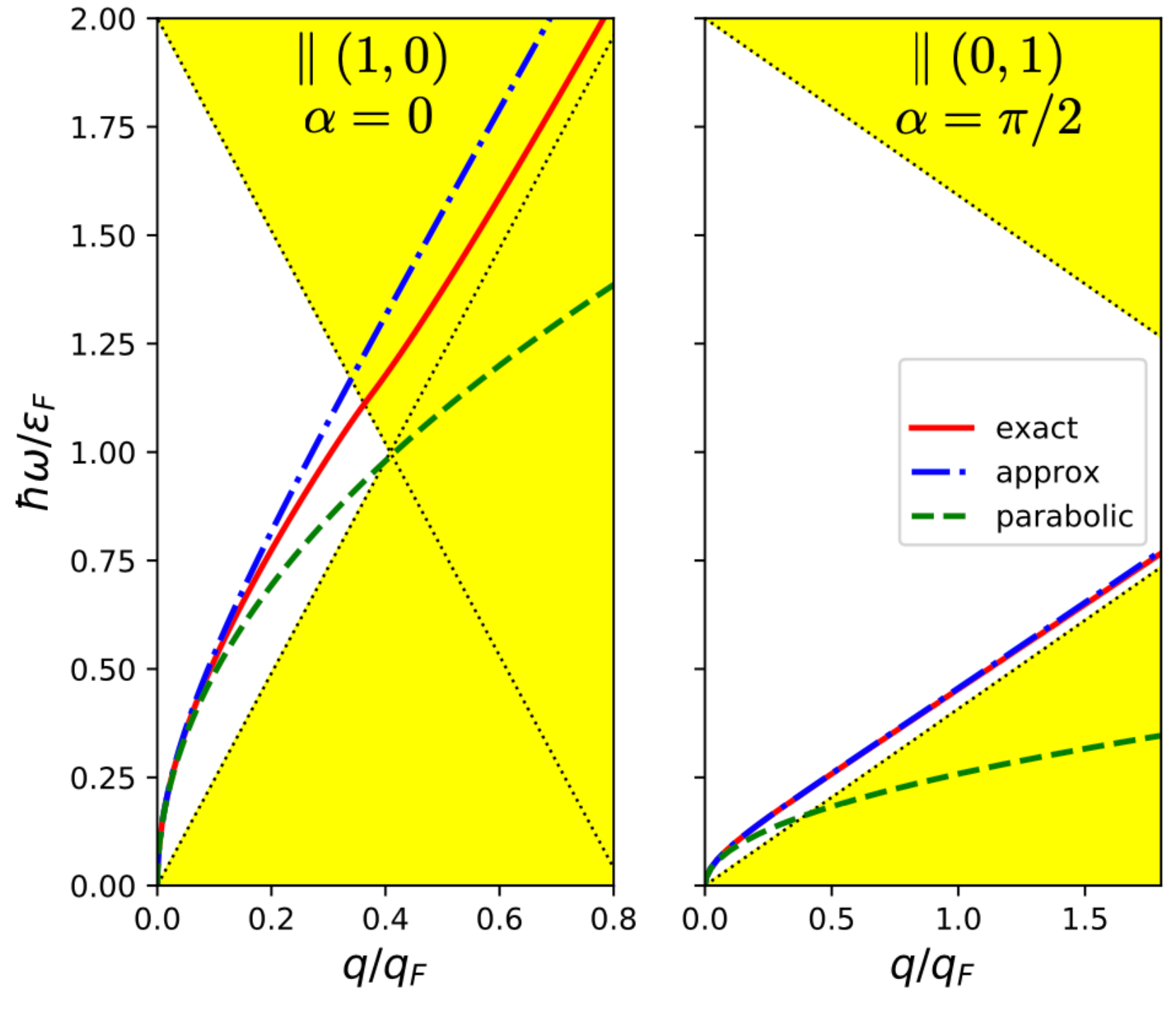}
\caption
{Plasmon dispersion for material parameters $\beta=0.4$ and $A=6.0$ for the two extremal plasmon
propagation directions
$\alpha=0$ (left) and $\alpha=\pi/2$ (right) using the exact expression or the two approximate
ones (see Figs.\ \ref{f1} and \ref{f2}). The continuum of electron-hole excitations
is indicated in yellow.
}
\label{f3}
\end{figure}
\begin{figure}
\vspace{1cm}
\centering
\includegraphics[width=0.9 \linewidth]{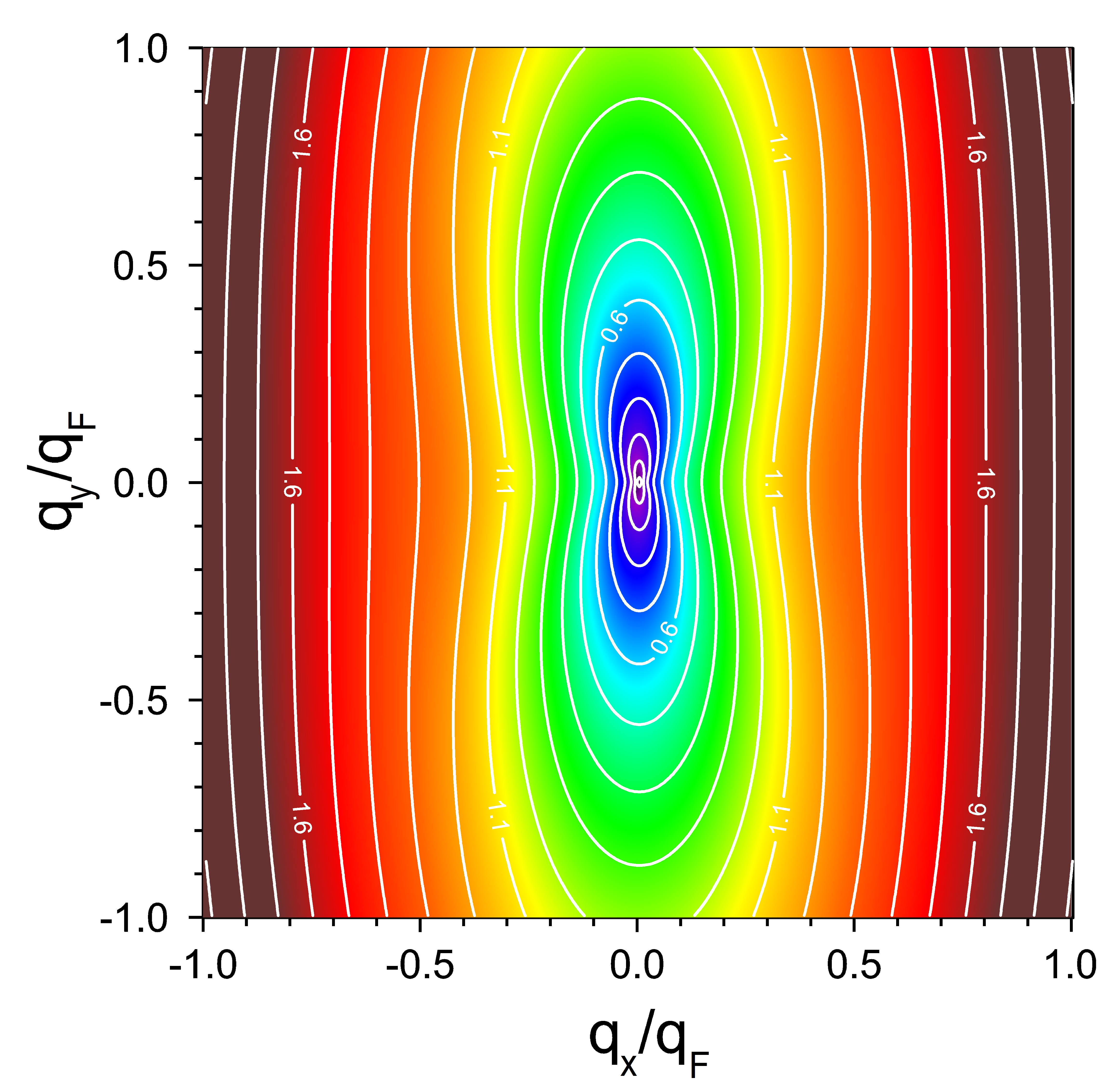}
\caption
{Contour plot of the anisotropic plasmon dispersion with lines of constant plasmon energy in the
$q_x$-$q_y$ plane for material parameters $\beta=2.0$ and $A=2.5$.}
\label{Fig}
\end{figure}

The exact plasmon dispersion together with that one resulting from the two approximations (\ref{eq11})
and (\ref{eq12}) is shown
in Figs.\ \ref{f2} and \ref{f3} for two different sets of material parameters. In all cases, we show the
two extremal directions $\alpha=0$ and
$\alpha=\pi/2$. The behavior is different for materials with $\beta$ larger than one and having a relatively small
anisotropy (represented in Fig.\ \ref{f2} for $\beta=2.0$ and $A=1.5$) from that one for $\beta$ being considerably
smaller than one and having a large anisotropy (Fig.\ \ref{f3} for $\beta=0.4$ and $A=6.0$). In Fig.\ \ref{f2} both
approximations represent relatively well the exact plasmon dispersion. The square-root dispersion never crosses the line
$\nu=\kappa$ and enters into the continuum of electron-hole excitations where it gets a final life-time by crossing
the upper line $\nu=2-\kappa$. That is different in Fig.\ \ref{f3}. There the square-root dispersion crosses the line
$\nu=\kappa$ which is especially visible for $\alpha=\pi/2$ in the right hand part of the figure. Just relying  on the parabolic
approximation would imply that the plasmon becomes damped above a critical value $\nu_c=\beta/\sqrt{A}$ which was
incorrectly inferred in Ref.\ \onlinecite{Raghu10} for Bi$_2$Se$_3$. In effect, due to the divergence of $\chi_0$ at
$\nu=\kappa$, the exact plasmon dispersion can never cross the line $\nu=\kappa$ such that the plasmon
remains undamped up to a critical $\nu_c$ of order one.
Lines of constant plasmon energy are shown in Fig. \ref{Fig}  for $\beta=2.0$ and $A=2.5$.
Clearly, they deviate strongly from simple ellipses which are expected for a tilted Dirac cone \cite{Jalali18} and show a
remarkable anisotropy which increases at small plasmon energies.

\section{Materials}

The material parameter $\beta$ can vary quite considerably in different Dirac systems. For graphene with $g=4$,
$\varepsilon_{rel}=2.4$, and $v=9 \times 10^5 ms^{-1}$, one finds $\beta=2.08$, exceeding $\beta=1$ considerably.

For Bi$_2$Se$_3$ a bulk dielectric constant perpendicular to the $c$-axis of $\varepsilon_{rel}\approx27$ was obtained in the first-principles calculations.\cite{nehaprb13} Experimentally determined $\varepsilon_{rel}\approx29$ was obtained using single crystals.\cite{Richter77,Stordeur92} Employing the latter value  together with
$v=5 \times 10^5 ms^{-1}$ and $g=1$ leads to $\beta=0.10$,
which compares well with the simulation of the measured plasmon
dispersion in Ref.\ \onlinecite{Politano15}.


Turning to anisotropic Dirac systems we have to distinguish two different cases, systems like HgS or Ag$_2$Te with a
preferred direction of plasmon propagation, or materials of the BaMnBi$_2$ class which preserve a fourfold rotation
symmetry axis at the surface despite the strong anisotropy of the Dirac cones. Our theory directly applies to the
first class of systems. The anisotropy factor was predicted to be 18 (HgS)\cite{Virot11} or
about 10 (Ag$_2$Te)\cite{Zhang11}. The $\beta$ parameter is more difficult to estimate due to the uncertain
knowledge about $\varepsilon_{rel}$. By comparison with Bi$_2$Se$_3$ the $\beta$ parameter can be
assumed to be smaller or close to 1. So one expects a scenario close to that reported in Fig.\ \ref{f3}, with an even higher anisotropy
factor $A$.

For the other class of anisotropic Dirac cones with conserved 4 fold rotation symmetry, there are all together
4 anisotropic Dirac cones being pairwise
perpendicular to each other. Therefore, one obtains four contributions to $\chi_0$:
\begin{equation}
\chi_0 = \frac{g \varepsilon_F}{4 \pi} \left( \frac{ {{\bf Q}_1}^2}{\omega^2}
+ \frac{ {{\bf Q}_2}^2}{\omega^2}  \right) \; ,
\end{equation}
where the preferred direction of one cone ${{\bf Q}_1}^2=q^2(A\cos^2 \alpha + \sin^2 \alpha/A)$ is perpendicular
to that one of the other cone ${{\bf Q}_2}^2=q^2(A\sin^2 \alpha + \cos^2 \alpha/A)$ and $g=2$.
We see that the anisotropy disappears in the leading order and remains only in higher orders. The anisotropy
is expected to be much smaller than in the other class of anisotropic TI's and to appear only for larger values of $q$
as is quite usual in many realistic materials.

%
%

\section{Discussion and conclusions}

We have shown how the well-known square-root dispersion for 2D Dirac plasmons can be generalized to the
anisotropic case. Interestingly, the entire material class of anisotropic Dirac systems can be described by
just two material parameters $\beta$ and $A$. For materials with small values of $\beta$ the square-root dispersion
applies only for very small frequencies and has to be replaced by a more exact one close to the continuum of
electron-hole excitations. Materials with high anisotropy factor $A$ show strongly anisotropic plasmon
excitations in the entire energy range  up to
very small frequencies.
Controlling either of the material parameters opens the pathway to engineer and customize 2D Dirac systems for plasmonics.
In particular, for high anisotropies, plasmon wave guides may be constructed.\cite{maaln18,mashim20}
It should be mentioned, that anisotropic
low-energy plasmons can also be realized in other 2D systems with a non-linear band dispersion like
phosphorene,\cite{loroprl14,lagujap15,ghthprb17,savaprb17,lemaap18,wazhaom20} borophene,\cite{hushjacs17,lihuprl20,delioe20} and
MoS$_2$.\cite{toasjpcm17} However, the amount of anisotropy is not so pronounced there as we predict here.


Verifying the predicted anisotropy of the plasmon dispersion requires measurements at the surface of anisotropic
TI's. One interesting candidate system is Ag$_2$Te for which the anisotropic Dirac cone was experimentally verified.
A useful technique to measure the plasmon dispersion at the surface of a TI is
electron energy loss spectroscopy (EELS) in reflection geometry. Also optical measurements are possible that
require periodic structure modifications, for instance surface grating.

{\em Acknowledgements} - R.H. thanks M. Knupfer, S.-L. Drechsler, and M. Richter for very helpful discussions.
V.M.S. acknowledges financial support from the Spanish Ministry Science and Innovation (Grant No. PID2019-105488GB-I00).
J.v.d.B. acknowledges financial support from the German Research Foundation (Deutsche Forschungsgemeinschaft, DFG) via SFB1143 Project No. A5 and under Germanys Excellence Strategy through the W\"urzburg-Dresden Cluster of Excellence on Complexity and Topology in Quantum Matter ct.qmat (EXC 2147, Project No. 390858490).


\begin{thebibliography}{}

\bibitem{Geim07}
A. K. Geim and K. S. Novoselov, Nature Mater.\ {\bf 6}, 183 (2007).
\bibitem{Kane05}
C. L. Kane and E. J. Mele, Phys.\ Rev.\ Lett.\ {\bf 95}, 146802 (2005).
\bibitem{Grigorenko12}
A.N. Grigorenko, M. Polini, and K. S. Novoselov,
Nature \ Photon.\ {\bf 6}, 749 (2012).
\bibitem{kochnl11} F. H. L. Koppens, D. E. Chang, and F. J. Garcia de Abajo, Nano Lett. {\bf 11}, 3370 (2011).
\bibitem{diornn13} P. Di Pietro, M. Ortolani, O. Limaj, A. Di Gaspare, V. Giliberti, F. Giorgianni, M. Brahlek, N. Bansal, N. Koirala, S. Oh, P. Calvani, and S. Lupi, Nat. Nanotechnol. {\bf 8}, 556 (2013).
\bibitem{gaap14} F. J. Garcia de Abajo, ACS Photonics {\bf 1}, 135 (2014).
\bibitem{Virot11}
F. Virot, R. Hayn, M. Richter, and J. van den Brink, Phys.\ Rev.\ Lett.\ {\bf 106}, 236806 (2011).
\bibitem{Zhang11}
W. Zhang, R. Yu, W. Feng, Y. Yao, H. Weng, X. Dai, and Z. Fang,
Phys.\ Rev.\ Lett.\ {\bf 106}, 156808 (2011).
\bibitem{gievsr14} Q. D. Gibson, D. Evtushinsky, A. N. Yaresko, V. B. Zabolotnyy, M. N. Ali, M. K. Fuccillo, J. van den Brink, B. B\"uchner, R. J. Cava, and S. V. Borisenko, Sci. Rep. {\bf 4}, 5168 (2014).
\bibitem{fewasr14} Y. Feng, Z. J. Wang, C. Y. Chen, Y. G. Shi, Z. J. Xie, H. A. Yi, A. J. Liang, S. L. He, J. F. He, Y. Y. Peng, X. Liu, Y. Liu, L. Zhao, G. D. Liu, X. O. Dong, J. Zhang, C. T. Chen, Z. A. Xu, X. Dai, Z. Fang, and X. J. Zhou, Sci. Rep. {\bf 4}, 5385 (2014).
\bibitem{Ryu18} H.Ryu, S.Y. Park, L.Li, W. Ren, J.B. Neaton, C. Petrovic, C. Hwang, and S.-K. Mo, Sci. Rep. {\bf 8}, 15322  (2018).
\bibitem{himiepl09} A. Hill, S. A. Mikhailov, and K. Ziegler, EPL {\bf 87}, 27005 (2009).
\bibitem{gayussc11} Y. Gao and Z. Yuan, Solid State Commun. {\bf 151}, 1009 (2011).
\bibitem{denoprb13} V. Despoja, D. Novko, K. Dekani\'c, M. \v{S}unji\'c, and L. Maru\v{s}i\'c, Phys. Rev. B {\bf 87}, 075447 (2013).
\bibitem{pisinjp14} M. Pisarra, A. Sindona, P. Riccardi, V. M. Silkin, and J. M. Pitarke, New J. Phys. {\bf 16}, 083003 (2014).
\bibitem{Choi10}
S.-M. Choi, S.-H. Jhi, and Y.-W. Son, Phys.\ Rev.\ B {\bf 81}, 081407 (2010).
\bibitem{Tajima06} N. Tajima, S. Sugawara, M. Tamura, Y. Nishio, and K. Kajita, J.\ Phys.\ Soc.\ Jpn.\ {\bf 75}, 051010 (2006).
\bibitem{Feng17} B. Feng, O. Sugino, R.-Y. Liu, J. Zhang, R. Yukawa, M.
Kawamura, T. Iimori, H. Kim, Y. Hasegawa, H. Li, L. Chen, K. Wu, H. Kumigashira, F. Komori, T.-C. Chiang, S. Meng, and I. Matsuda,
Phys.\ Rev.\ Lett.\ {\bf 118}, 096401 (2017).
\bibitem{Nishine10}
T. Nishine, A. Kobayashi, and Y. Suzumura, J.\ Phys.\ Soc.\ Jpn.\ {\bf 79}, 114715 (2010).
\bibitem{Sadhukhan17}
K. Sadhukhan and A. Agarwal, Phys.\ Rev.\ B {\bf 96}, 035410 (2017).
%
\bibitem{Sari14}
J. S\'ari, C. T\"oke, and M.O. Goerbig, Phys.\ Rev.\ B {\bf 90}, 155446 (2014).
%
\bibitem{Jalali18}
Z. Jalali-Mola and S. A. Jafari, Phys.\ Rev.\ B {\bf 98}, 195415 (2018).
%
%
\bibitem{Wunsch06}
B. Wunsch, T. Stauber, F. Sols, and F. Guinea, New J.\ Phys.\ {\bf 8}, 318 (2006).
%
\bibitem{Hwang07}
E. H. Hwang and S. Das Sarma, Phys.\ Rev.\ B {\bf 75}, 205418 (2007).
%
\bibitem{Principi09}
A. Principi, M. Polini, and G. Vignale, Phys.\ Rev.\ B {\bf 80}, 075418 (2009).
%
\bibitem{Raghu10}
S. Raghu, S. B. Chung, X.-L. Qi, and S.-C. Zhang, Phys.\ Rev.\ Lett.\ {\bf 104}, 116401 (2010).
%
%
\bibitem{nehaprb13} I. A. Nechaev, R. C. Hatch, M. Bianchi, D. Guan, C. Friedrich, I. Aguilera, J. L. Mi, B. B. Iversen, S. Bl\"ugel, Ph. Hofmann, and E. V. Chulkov, Phys. Rev. B {\bf 87}, 121111 (2013).
%
\bibitem{Richter77}
W. Richter, H. K\"ohler, and C. R. Becker, Phys. Stat. Sol. (b) {\bf 84}, 619 (1977).
%
\bibitem{Stordeur92}
M. Stordeur, K. K. Ketavonc, A. Priemuth, H. Sobotta, and V. Riede, Phys. Stat. Sol. (b) {\bf 169}, 505 (1992).
%
\bibitem{Politano15}
A. Politano, V. M. Silkin, I. A. Nechaev, M. S. Vitiello, L. Viti, Z. S. Aliev, M. B. Babanly, G. Chiarello,
P. M. Echenique, and E. V. Chulkov,  Phys.\ Rev.\ Lett.\  {\bf 115}, 216802 (2015).
%
\bibitem{maaln18} W.-L. Ma, P. Alonso-Gonz\'{a}lez, S.-J. Li, A. Y. Nikitin, J. Yuan, J. Mart\'{\i}n-S\'{a}nchez, J. Taboada-Guti\'{e}rrez, I. Amenabar, P.-N. Li, S. V\'{e}lez, C. Tollan, Z.-G. Dai, Y.-P. Zhang, S. Sriram, K. Kalantar-Zadeh, S.-T. Lee, R. Hillenbrand,  and Q.-L. Bao, Nature {\bf 562}, 557 (2018).
%
\bibitem{mashim20} W.-L. Ma, B. Shabbir, Q.-D. Ou, Y.-M. Dong, H.-Y. Chen, P.-N. Li, X.-L. Zhang, Y.-R. Lu, and Q.-L. Bao, InfoMat. {\bf 2}, 777 (2020).
%
\bibitem{loroprl14} T. Low, R. Rold\'an, H. Wang, F. Xia, P. Avouris, L. M. Moreno, and F. Guinea, Phys. Rev. Lett. {\bf 113}, 106802 (2014).
%
\bibitem{lagujap15} R.-T. Lam and J. Guo, J. Appl. Phys. {\bf 117}, 113105 (2015).
%
\bibitem{ghthprb17} B. Ghosh, P. Kumar, A. Thakur, Y. S. Chauhan, S. Bhowmick, and A. Agarwal, Phys. Rev. B {\bf 96}, 035422 (2017).
%
\bibitem{savaprb17} S. Saberi-Pouya, T. Vazifehshenas, T. Salavati-fard, and M. Farmanbar, Phys. Rev. B {\bf 96}, 115402 (2017).
%
\bibitem{lemaap18} I.-H. Lee, L. Martin-Moreno, D. A. Mohr, K. Khaliji, T. Low, and S.-H. Oh, ACS Photonics {\bf 5}, 2208 (2018).
%
\bibitem{wazhaom20} C. Wang, G. W. Zhang, S. Y. Huang, Y. G. Xie, and H. Yan, Adv. Opt. Mater. {\bf 8}, 1900996 (2020).
%
\bibitem{lihuprl20} C. Lian, S.-Q. Hu, J. Zhang, C. Cheng, Z. Yuan, S. Gao, and S. Meng, Phys. Rev. Lett. {\bf 125}, 116802 (2020).
%
\bibitem{delioe20} S. A. Dereshgi, Z. Z. Liu, and K. Aydin, Opt. Exp. {\bf 28}, 16725 (2020).
%
\bibitem{hushjacs17} Y. Huang, S. N. Shikodkar, and B. I. Yakobson, J. Amer. Chem. Soc. {\bf 139}, 17181 (2017).
%
\bibitem{toasjpcm17} Z. Torbatian and R. Asgari, J. Phys.: Condens. Matter {\bf 29}, 465701 (2017).


\end{thebibliography}
\end{document}